\documentclass[11pt]{article}
\usepackage[latin1]{inputenc}
\usepackage[english]{babel}
\usepackage[namelimits]{amsmath}
\usepackage{authblk}
\usepackage{amssymb}
\usepackage{amsmath}
\usepackage{amsthm}

\begin{document}
\title{{\bf Building a linear equation of state for trapped gravitons from finite size effects and the Schwarzschild black hole case}}
\author[1,2]{S. Viaggiu\thanks{viaggiu@axp.mat.uniroma2.it and s.viaggiu@unimarconi.it}}
\affil[1]{Dipartimento di Matematica, Universit\`a di Roma ``Tor Vergata'', Via della Ricerca Scientifica, 1, I-00133 Roma, Italy.}
\affil[2]{INFN, Sezione di Napoli, Complesso Universitario di Monte S. Angelo,
	Via Cintia Edificio 6, 80126 Napoli, Italy.}

\date{\today}\maketitle
\begin{abstract}
In this paper we continue the investigations present in \cite{1} and \cite{2}
concerning the spectrum of trapped gravitons in a spherical box, and
in particular inside a Schwarzschild black hole (BH). 
We explore the possibility that, due to finite size effects, the frequency of the radiation made of
trapped gravitons can be modified in such a way that a linear equation of state $PV=\gamma U$ for the pressure
$P$ and the internal energy $U$
arises. Firstly, we study the case with
$U\sim R$, where only fluids with $\gamma >-\frac{1}{3}$ are possible. If corrections  $\sim 1/R$ are added to $U$, for
$\gamma\in[0,\frac{1}{3}]$ we found no limitation on the allowed value for the areal radius of the trapped sphere $R$.
Moreover, for $\gamma>\frac{1}{3}$
we have a minimum allowed value for $R$ of the order of the Planck length $L_P$.  
Conversely, a fluid with 
$P<0$ can be obtained but with a maximum allowed value for $R$. With the added term looking like
$\sim 1/R$ to the BH internal energy $U$, the well known logarithmic corrections to the BH entropy naturally emerge
for any linear equation of state. The results of this paper suggest that finite size effects could modify the structure of 
graviton's radiation inside, showing a possible mechanism to transform radiation into dark energy.
 
\end{abstract}
PACS Number(s): 05.20.-y, 04.30.-w, 04.70.-s, 04.60.-m

\section{Introduction}

The discovery \cite{H} that BHs emit a practically thermal radiation and thus are equipped with a non-vanishing entropy
\cite{H,3} $S_{BH}$ given by $S_{BH}=K_B A_h/L_P^2$, where $K_B$ is the Boltzmann constant, $L_P$ the Planck length and 
$A_h$ the area of the event horizon, opened the doors to a possible deep understanding of quantum effects in a strong 
gravitational field with important still open issues. 
In particular, a complete statistical description of the BH entropy (see for example \cite{3,4,5,6,7,8,9,10,11,12,13,w1,w2,k1,k2,k3,k4,k5,k6,k7,k8,k9,k10,k11,k12,k13,k14}) 
is still lacking. In practically all the approaches present in the literature, one assumes the origin of the unknown BH degrees 
of freedom as reside on or just outside the event horizon, according to the holographic BH nature. Althought all these approaches are 
physically reasonable, it is not yet understood what kind of matter-energy could fill the interior of a a BH, i.e. the fate of any kind of matter-energy falling inside the event horizon.
This may look as a speculative issue, but this is not the case since the degrees of freedom leading to the BH entropy can well be
stored inside the horizon. In particular, we could invoke the interesting {\it BH complementarity conjecture}, suggested
by 't Hooft and Susskind (for a nice review see \cite{14} and references therein), affirming that
two copies of information are stored in the Hawking radiation and inside a BH. This implies that an (unlucky) observer inside a BH
should measure an entropy $S_{BH}=K_B A_h/L_P^2$. To this purpose, in \cite{1} we have suggested that the interior of a BH can be composed
of trapped gravitons forming a radiation field. Since the BH solution is a vacuum solution of the Einstein's equations, this
seems the most simple assumption. In fact any other matter-fluid should have a non vanishing energy-momentum tensor and this struggle
with the vacuun nature of the BH solution.
We have shown that the usual BH entropy formula can be obtained with a temperature
$T\simeq 2T_{BH}$, supporting a radiation field made of gravitons inside. There, the first law can be written inside the event horizon as
$TdS_{BH}=dU+PdV$, where $U=Mc^2$, $M$ is the ADM mass, $V$ the thermodynamic volume and $P$ the pressure with 
$PV=U/3$ and $T\simeq 2T_{BH}$. Hence, our model incorporate in a simple and natural way a pressure term, that is missing in the 
ordinary treatment. In some sense, our approach is similar to the John Wheeler's old geon
\footnote{I would like to thank Lee Smolin which suggested to me the relation between my approach and the Wheeler's one.}
construction \cite{w}, where a confinement of electromagnetic radiation is considered. In our approach the electromagnetic
radiation is substituted by gravitons.

In the paper \cite{2} we have analyzed the possible origin of corrections to the BH entropy formula by adopting suitable modifications of
the internal energy $U$ dictated by quantum phenomena, but retaining the radiation character of the trapped gravitons by fixing
$\gamma=1/3$.
A possible interesting and intriguing development of the model present in \cite{1} 
and further analyzed in \cite{2},
considers the possibility
to obtain a linear equation of state, i.e. $PV=\gamma U,\;\gamma\in\mathbf{R}$, for trapped gravitons inside the event horizon caused 
by possible finite size effects induced by a finite radius $R$
and then to study 
possible limitations to the areal radius $R$ of the event horizon. The answer present in this paer is affirmative.

In section 2 we present and derive the formalism and the key equations for our purposes. In section 3 we present the main formulas in a compact form as a theorem.
In section 4 we study the general case for a classical BH, while in section 5 we analyse the case with $U\sim R^{n+1}$
with $n\in\mathbf{N}$. In section 6 we investigate the case where strong quantum effects modify the expression for the internal energy
$U(R)$ together with the intriguing and
important case of dark-energy equation of state. 
Finally, section 7 collects conclusions and final remarks. 

\section{Preliminaries}

To obtain a suitable formula for trapped gravitons, in \cite{1} we considered a spherical gravitational wave traveling (perturbing)
a Minkowski spacetime. After imposing the Dirichelet boundary condition on the areal radius $R$ of the confining ball, the discrete spectrum 
so obtained is nothing else but:
\begin{equation}
{\omega}_{\ell n}\simeq\frac{c}{2R}\left(2+\ell+2n\right)\pi,\,\;\ell\geq 2,\;\;n\in\mathbb{N},
\label{1}
\end{equation}
where $\{\ell\}$ is the integer Legendre index with $\ell \geq 2$ for a gravitational wave, 
$n$ is an integer quantum number and $c$ denotes the speed of the light. As shown in \cite{1}
and \cite{2}, formula 
(\ref{1}) is a good approximation to explore the physical features of the statistical mechanics of trapped gravitons, with the index
$\ell$ labeling the species of trapped gravitons (quadrupolar, sextupolar $\cdots$) rather than a quantum number. 
After introducing
an ensemble of $N$ harmonic oscillators made of trapped gravitons, for the partition function
$Z_T^{(0)}=Z_g^{(0) N}$ we have:
\begin{equation}
Z_g^{(0)}=\sum_{\ell=2}^{\infty}\sum_{n=0}^{\infty} e^{-\beta\hbar{\omega}_{\ell n}}=
\frac{e^{-\left(\frac{2c\pi\beta\hbar}{R}\right)}}
{\left[1-e^{-\left(\frac{c\pi\beta\hbar}{2R}\right)}\right]}
\frac{1}{\left[1-e^{-\left(\frac{c\pi\beta\hbar}{R}\right)}\right]},
\label{2}
\end{equation}
where $\beta=1/(K_B T)$. 
For the entropy $S^{(0)}$, the internal energy $U^{(0)}$ and the pressure $P^{(0)}$, thanks to the (\ref{2}) we obtain
\begin{eqnarray}
& & S^{(0)}=-N K_B\left[\ln\left(1-e^{-\frac{X}{2}}\right)+\ln\left(1-e^{-X}\right)\right]+\label{3}\\
& & +\frac{c\pi\hbar N e^{-\frac{X}{2}}\left[1+3 e^{-\frac{X}{2}}\right]}{2TR\left(1-e^{-X}\right)},\;\;
X=\frac{c\pi\beta\hbar}{R},\nonumber\\
& & U^{(0)}=\frac{c\pi\hbar N}{2R\left[e^{\frac{\beta c\pi\hbar}{2R}}-1\right]}+\frac{c\pi\hbar N}{R\left[e^{\frac{\beta c\pi\hbar}{R}}-1\right]}.
\label{4}\\
& & P^{(0)}=\frac{c\pi N\hbar\left[e^{-\frac{X}{2}}+3e^{-X}\right]}{2R^2\left(1-e^{-X}\right)}\frac{dR}{dV},
\label{5}
\end{eqnarray}
where $V=\frac{4\pi R^3}{3}$ and  $dR/dV=3^{-1}{(3/(4\pi))}^{1/3} V^{-2/3}$. 
Note that, as customary, the zero point energy has been subtracted. It is trivial to verify that we have $P^{(0)}V=U^{(0)}/3$. Hence, the spectrum formula (\ref{1}) leads to a radiation field. For a BH we have $U^{(0)}(R)=c^4 R/(2G)=Mc^2$, with
$M$ the ADM mass of the BH. 
As shown in \cite{1}, after solving the (\ref{4}) for $N$ and substituing the so obtained expression in (\ref{3}), we obtain
\begin{eqnarray}
& & T=\alpha T_{BH}=\frac{\alpha c\hbar}{4\pi K_B R},\;\;\alpha\in (0,\infty),\label{6}\\
& & S=K_B Y(\alpha)\frac{A_h}{4L_P^2},\;\;A_h=4\pi R^2,\label{7}\\
& & Y(\alpha)=\frac{b}{\alpha\pi^2\left(3+e^{\frac{2\pi^2}{\alpha}}\right)}\nonumber\\
& & b=-\alpha e^{\frac{4\pi^2}{\alpha}}\ln\left(1-e^{-\frac{2\pi^2}{\alpha}}\right)-
\alpha e^{\frac{4\pi^2}{\alpha}}\ln\left(1-e^{-\frac{4\pi^2}{\alpha}}\right)+6\pi^2+\nonumber\\
& & +2\pi^2 e^{\frac{2\pi^2}{\alpha}}+\alpha \ln\left(1-e^{-\frac{2\pi^2}{\alpha}}\right)+
\alpha\ln\left(1-e^{-\frac{4\pi^2}{\alpha}}\right). 
\end{eqnarray}
By setting $\alpha\simeq 2.2$, we arrive to the BH entropy formula.

\section{Building linear equations of state: mathematical formulation}

In \cite{1} and \cite{2} we have considered gravitons as trapped particles representing a radiation field. A fundamental physical ingredient in \cite{1}-\cite{2} is the discrete spectrum (\ref{1}) for trapped gravitons. As shown in \cite{1}, the spectrum 
(\ref{1}) does imply the equation of state $PV=U/3$, i.e. a radiation field. In \cite{15} we have
shown that finite size effects induced by the apparent horizon of a Friedmann flat universe could imply a kind of interaction between gravtitons that are supposing to compose the 
cosmological constant. These finite size effects could also modify, as an example, the thermodynamic of a Friedmann universe
\cite{17}. Following these results, and in particular the one present in \cite{15}, it
can be 
supposed that some interaction can take place between gravitons inside the trapped sphere and as a result, the equation of state of the trapped gravitons will have modified result. As an example, we may suppose that some quantum field $\Phi(R)$, generated by long range quantum fluctuations, is capable to change the graviton's equation of state. Hence, in this paper we do not enter in reasonings concerning the 
physical origin of $\Phi$. Rather, we address the following interesting problem. Suppose to have a massless radiation field, in our case made of gravitons with the discrete spectrum (\ref{1}): how it should be changed the spectrum (\ref{1}) in order to obtain a 
linear equation of state $PV=\gamma U$ (with $\gamma\in\mathbf{R}$ and $V=4\pi R^3/3$)? This is an intriguing problem that can have important applications also in cosmology. To be more quantitative and without loss of generality, suppose to start with a physical situation where gravitons are trapped with angular frequency ${\omega}_0$ given by the (\ref{1}). We wonder if there exists
a frequency ${\omega}={\omega}_0+\Phi(R)$ such that a linear equation of state arises for trapped gravitons. To this purpose,
we may reasonable suppose, without loss of generality, 
that the frequency of a single graviton can change by an amount depending on the collective presence of the other gravitons,
as depicted in \cite{15}, i.e. $\Phi(R,N)=\frac{\phi(R)}{N}$ where $N$ is the graviton's number. The following proposition 
holds (where commas "," denotes partial derivative):

\noindent{\bf Proposition:} Let ${\omega}_0$, given by the (\ref{1}), denote the angular frequency of $N$ trapped gravitons. The so trapped
gravitons 
with angular frequency $\omega={\omega}_0+\frac{\phi(R)}{N}$ have a linear equation of state
$PV=\gamma U$ if and only if
there exists a differentiable function $\frac{\phi(R)}{N}$, satisfying the following equation 
\begin{equation}
\hbar\left[R\;{\phi}_{,R}(R)+\phi(R)\right]=U(R)(1-3\gamma), \label{10}
\end{equation}
together with the condition
\begin{equation}
U-\hbar\;\phi(R) > 0.
\label{11}
\end{equation}
\begin{proof}
To start with, following the notation of equations (\ref{2})-(\ref{7}), we have
\begin{equation}
Z_{T} = e^{-\beta\hbar\phi}\;Z_T^{(0)}.
\label{12}
\end{equation}
From (\ref{12}), with the usual relation $U=-{\ln(Z_T)}_{,\beta}$, we obtain
\begin{equation}
U=U^{(0)}+\hbar\;\phi(R).
\label{13}
\end{equation}	
Since from the (\ref{4}) we have $U_0>0$, condition (\ref{11}) follows. For the free energy we have
$F_T=-K_B T\ln(Z_T)=F_T^{(0)}+\hbar\;\phi(R)$. Moreover
\begin{equation}
{F_T}_{,V}=\hbar\;{\phi}_{,R}\;R_{,V}+R_{,V}\;{F_T^{(0)}}_{,R}=-P,
\label{14}
\end{equation}
with $R_{,V}\;{F_T^{(0)}}_{,R}=-P^{(0)}$ and $P^{(0)}V=\frac{U^{(0)}}{3}$. Hence, from (\ref{14}) we get
\begin{equation}
\hbar\frac{R}{3V}\;{\phi}_{,R}-\frac{U^{(0)}}{3V}=-P.	
\label{15}
\end{equation}	
Finally, after using the (\ref{13}) and imposing $PV=\gamma U$, from (\ref{15}) we obtain the equation (\ref{10}). 
\end{proof}
On general grounds, we could expect the added frequency $\frac{\phi}{N}$ as vanishing for
$N\rightarrow\infty$ ($R\rightarrow\infty$), according to the possible quantum nature of $\phi(R)$ due to finite size effects,
provided that a physically viable expression for $U$ is given. In the following sections we show that this is indeed the case. 

As well known, the general solution of the ordinary first order differential equation (\ref{10}) is provided by the sum of the general solution of the homogeneous equation ($U=0$) with a particular solution with $U\neq 0$. It is trivial to see that the solution 
${\phi}_0$ of the homogeneous part of the (\ref{10}) looks like ${\phi}_0\sim\frac{1}{R}$. Since the spectrum (\ref{10}) 
is proportional to
$\frac{1}{R}$, the homogeneous solution of the (\ref{10}) gives a contribution for the zero point energy \cite{1} and can thus, 
as customary, subtracted and as a result plays no role. This choice is further motivated by the fact that equation (\ref{10}) becomes 
homogeneous for a radiation field with $\gamma=\frac{1}{3}$ and thus we impose the condition that for a radiation field the spectrum
(\ref{1}) is regained, i.e. we set $\phi=0$ as the solution of the homogeneous part of (\ref{10}). Also note that the condition
(\ref{11}), for a given choice of $U(R)$, could imply an inequality for the areal radius $R$. Hence, as shown in the following sections, generally we expect some limitations to the range of $R$, it crucially depending on the value of $\gamma$.

\section{General case for a classical Schwarzschild BH}

The first case we study is provided by the Schwarzschild BH case \cite{1}-\cite{2} where $U=\frac{c^4 R}{2G}$.
After inserting the above expression for $U$ in (\ref{10}), we obtain the solution for $\phi$ given by:
\begin{equation}
\phi(R)=\frac{c}{4L_P^2}R(1-3\gamma).
\label{16}
\end{equation}
Condition (\ref{11}) becomes
\begin{equation}
\frac{c^4 R}{2G}-\frac{c\hbar R}{4L_P^2}(1-3\gamma)>0,
\label{17}
\end{equation}
with the solution $\gamma >-\frac{1}{3}$. As a consequence, no limitation on $R$ arises but only solutions with $\gamma >-\frac{1}{3}$
are allowed
\footnote{Note that solutions with $\gamma=-\frac{1}{3}$ are allowed only in the limit for $T\rightarrow 0$.}. Since 
 $3P+\frac{U}{V}>0$ does imply a positive active gravitational mass in general relativity, we obtain the 
rather physically reasonable result that with the classical expression
for $U$ only configurations with positive active gravitational mass are allowed. In particular, we obtain that a dark energy-like
equation of state with $\gamma=-1$ is incompatible with the classical expression for $U$.
To allow configurations with negative active gravitational mass for a BH quantum, modifcations to the classical behavior $U(R)\sim R$ must be considered (see section 6). Concerning the internal temperature $T_{iBH}$ of the BH, we must have, as shown in \cite{2},
$T_{iBH}=\alpha T_{BH}=\frac{\alpha c\hbar}{4\pi K_B R},\;\;\alpha\in (0,\infty)$. With this choice for
$T_{iBH}$ we have the usual expression $S_{BH}=\frac{A}{4L_P^2}$, but
only for dust's gravitons with $\alpha=1$ we have $T_{iBH}=T_{BH}$.\\ 
Summarizing, also supposing gravitons filling the interior of a
BH with a linear equation of state with positive active gravitational mass, the usual holographic relation $S_{BH}\sim A$ is regained.

\section{Case with $U\sim R^{n+1}$, $n\in\mathbf{N}$}

In the spirit of the study present in \cite{2}, we can also consider expressions for $U(R)$ that are not suitable for a BH case.
To this purpose, we set $U=\frac{C_n c^4}{2G}\frac{R^{n+1}}{L_P^n}$, with $C_n\in{\mathbf{R}}^+$ and
$n\in\mathbf{N}$. After solving equation (\ref{10}) for $\phi(R)$ we obtain
\begin{equation}
\phi(R)=\frac{c C_n R^{n+1}}{2(n+2)L_P^{n}}(1-3\gamma).
\label{18}
\end{equation}
With the (\ref{18}), condition (\ref{11}) it gives
\begin{equation}
\gamma > \frac{-n-1}{3}.
\label{19}
\end{equation}
Setting $n=0$, we obtain the solution (\ref{16}). Also in this case, limitations arise only for $\gamma$ and not for $R$, but with dark
energy configurations available for $n>2$.

\section{Quantum corrections to the internal energy $U(R)$}

Corrections to the BH entropy are expected to hold in any consistent quantum gravity theory 
(See for example \cite{14}, \cite{2} and references therein). 
Many physical and mathematical arguments
(see for example \cite{20} for a recent review and references therein) lead to a modified expression for 
$S_{BH}$ given by
\begin{equation}
S_{BH}=\frac{A}{4L_P^2}-a_0\ln\left(\frac{A}{4L_p^2}\right)+higher\:\:order\;\;corrections\;.
\label{20}
\end{equation}
For our purpose, the starting point is a suitable modified physically motivated expression for $U$. Such an expression can be 
well motivated in the context of a quantum spacetime \cite{21}-\cite{22}. There, physically motivated spacetime uncertainty 
relations can be obtained \footnote{Physically motivated uncertainty relations for a quantum spacetime can be found in
	\cite{21} in Newtonian approximation, improved in \cite{23} by using the isoperimetric Penrose's inequality and further generalized in
	\cite{24} in a Friedmann flat universe}. For a spherical configuration (see \cite{25} and the discussion in \cite{23}),
they reduce to
\begin{equation}
\Delta R\geq s L_P,\;\;\;c\Delta t\Delta R\geq s^2L_P^2,\;\;s\sim 1,
\label{21}
\end{equation}
where $\Delta$ denotes the uncertainty in a generic quantum state $\omega$.
If spherical symmetry is assumed with $c\Delta t\sim \Delta R$, then
a minimal uncertainty for $\Delta R$ arises. Moreover, since from ordinary quantum mecanics we have
$\Delta E\Delta t\geq\frac{\hbar}{2}$, by considering spherical states 
minimizing the uncertainty relations, we have 
$\Delta E\sim 1/R$. As a consequence, the expression for $U$ above justified becomes
\begin{equation}
U=\frac{c^4}{2G}R+\frac{C_1 c^4 L_P^2}{2GR},\;\;\;\;\{C_1\}\in{\mathbf{R}}^+.
\label{22}
\end{equation}
With the (\ref{22}), the solution of the (\ref{10}) is given by
\begin{equation}
\phi(R)=\frac{c R}{4 L_P^2} (1-3\gamma)+\frac{c\;C_1\;(1-3\gamma)}{2R}\ln\left(\frac{R}{sL_P}\right)
\label{23}
\end{equation}
First of all, it is worth to be noticed that with the term looking like $\ln(R)/R$, the logarithmic corrections given by the
(\ref{20}) do arise as a consequence of the term proportional to $1/R$ in (\ref{22}) and the request to have a linear equation of 
state for the trapped gravitons. This is because entropy (\ref{3}), independently on the equation of state, is proportional to 
$N$. From formula (\ref{4}) and (\ref{13}) we see that $N\sim R(U-\hbar\phi)$ and thus formula (\ref{20}) emerges.
This is certainly an intriguing fact. In the paper \cite{2} we have shown as logarithmic corrections can arise by considering the gravitons forming a pure radiation. Also by assuming a mechanism transforming a radiation of massless gravitons to massless ones but with a linear equation of state, logarithmic corrections do emerge, provided that correction to $U(R)$ looking like 
$1/R$ is considered caused by possible quantum fluctuations. This implies that our 'phenomenological'
approach, under reasonable hypothesis, is capable to capture the well wnown logarithmic corrections to the BH entropy independentely on the equation of state satisfied by gravitons.

Concerning the condition (\ref{11}), from the (\ref{23}) and after posing $W=\frac{R}{L_P}$, we get
\begin{equation}
W+\frac{C_1}{W}-(1-3\gamma)\left[\frac{W}{2}+\frac{C_1}{W}\ln\left(\frac{W}{s}\right)\right] > 0.
\label{24}
\end{equation}
First of all note that, thanks to the non-commutative arguments of equation (\ref{21}), 
the constant $s$ can be chosen of the order of unity. 
Also the positive constant $C_1$ \cite{2} can reasonable be fixed of the order of unity.\\
From a study of the (\ref{24}), we see that for $\gamma=\frac{1}{3}$, no condition arises, as expected for a radiation field
\cite{2}. More generally for $\gamma\in[0,\frac{1}{3}]$ no restriction to $R$ emerges. Conversely,
for $\gamma>\frac{1}{3}$ a minimum radius $R_{m}$ arises, but not a maximum one.
Finally, for $\gamma <0$ a maximum allowed radius $R_{M}$ arises, but not a minimum radius. These results are in agreement with physical intuition. In fact, for a fluid becoming more and more stiff (with $\gamma>\frac{1}{3}$) we expect that a huge core does appear taking 
the configuration no further compressible. In this condition, the BH evaporation can stop. This mechanism is not alternative to the one depicted in \cite{2}, where the term $\sim 1/R$ in $U(R)$ permit the end of the evaporation at some radius of the order of the
Planck length. For 'soft' fluid with $\gamma\in[0,\frac{1}{3}]$ no limitations arise. Conversely, fluid with 
$\gamma <0$ all have a negative pressure 
and thus we expect that configurations with a large radius cannot exist\footnote{This is expected in the static case, but in a cosmological context, as a Friedmann universe equipped with a cosmological constant, a stable configuration can exist.}. 
For numerical examples illustrating these results see the next two sections.

\subsection{Dark energy}
In this section, for its natural interest, we numerically analyze in more detail the case of the dark energy with $\gamma=-1$, in particular the existence condition (\ref{24}). First of all, we note that the value of $R_{M}(C_1,s)$ slowly run as a function of $C_1$.
By fixing $s=1$, we have, for $C_1=1$, $R_{M}=L_P$, for $C_1=2$, $R_{M}\simeq 1.1 L_P$, while for $C_1=10$ we have
$R_M\simeq 1.24 L_P$.\\
Also the behavior of $R_M$, obtained by varying $s$, is practically of the order of Planckian scales. As an example, after fixing
$C_1=1$, we get $R_M(s=10)\simeq 3.1$; $R_M(s=10^2)\simeq 3.8 L_P$; $R_M(s=10^3)\simeq 4.7 L_P$; $R_M(s=10^{20})\simeq 13.2 L_P$;
$R_M(s=10^{10^2})\simeq 30 L_P$.\\
The numerical examples above clearly show that BH filled with gravitons with a dark energy equation of state are only possible with 
microscopic Planckian sizes. As shown in \cite{2}, the term $\frac{1}{R}$ in $U(R)$ stops the BH evaporation process to a scale of the order
of the Planck length. We thus can suggest that BH reaching such a minimum radius could survive in a state with a dark energy equation of state, i.e. composed of gravitons with negative (repulsive) active gravitational mass. If we consider an ensemble of BH remnants
so obtained, a gas of 'dark' BH remnants can be built. This suggested mechanism could thus explain the nature of the dark energy.

\subsection{Other cases}

As stated above, models with $\gamma\in[0, \frac{1}{3}]$ have no restriction for $R$, while for $\gamma >\frac{1}{3}$ a minimum allowed value for $R$ is present.\\ 
As a numerical example, we consider the stiff case with $\gamma=1$. After fixing thereafter $s=1$, we have for
$R_m(C_1)$: $R_m(1)\simeq 0.5 L_P$; $R_m(5)\simeq 0.57$; $R_m(10)\simeq 0.59$. As a consequnce, a minimum is obtained
of the order of approximatively one half the Planck length.\\
For $\gamma<0$ a maximum allowed radius does appear. For $\gamma=-\frac{1}{3}$ we have $R_M(1)=\sqrt{e}L_P$, while for
$\gamma=-\frac{2}{3}$ we found $R_M(1)\simeq 1.1 L_P$. Finally, for $\gamma=-\frac{3}{2}$ (phantom fluid) we get
$R_M\simeq 0.9 L_P$.\\
From these numerical examples, it emerges that, when present, the limitations for $R$, min. or max, are of Planckian size.
According to the results of this section, we could suppose to
 have microscopic BH remnants composed with negative active gravitational mass (dark energy) and the ones with positive active gravitational mass that could be a possible candidate for dark matter \cite{22}.

\section{Conclusions and final remarks}

In this paper we continued the investigations present in \cite{1} and \cite{2}. In particular, we address the issue concernig the modification to be added to (\ref{1}) in order to have a linear equation of state for trapped gravitons. Physically speaking,
we may suppose that, for example, finite size effects and quantum effects could modify the equation of state of the gravitons inside.
In this regard, we found a simple but intriguing equation, namely equation (\ref{10}), that permit us to obtain the modification to the
(\ref{1}) suitable to obtain a linear equation of state. The condition (\ref{11}) poses possible restrictions on the values of $R$, it
depending on $\gamma$. For $\gamma >\frac{1}{3}$ a minimum radius does appear, while for $\gamma < 0$ a maximum one arises.
In particular, dark energy-like configurations are possible only with Planckian dimension, while configurations with 
$\gamma >\frac{1}{3}$ are possible practically up to Planckian dimensions. According with these reasonong, if the machanism
suggested in this paper is correct, macroscopic BH filled with massless gravitons with an equation of state different from the
radiation one, are possible only for $\gamma\geq 0$. More exotic configurations with $\gamma <0$ take place only at the 
Planck length, where quantum gravity effects are expected to be dominant.\\
The existence of possible BH remnants at the Planck length, where evaporation can stop \cite{2}, with negative 
active gravitational mass, open the possibility to represent dark-energy as filled with microscopic BH.

As an important final consideration, it can be shown that the proposition of section 3 can be generalized to any massless 
excitation. This open the possibility to generalize our results also for non-trapped or marginally trapped massless excitations, with potential applications in a cosmological context.

\end{document}